# Collective modes, Yb-valence instability and metal-insulator transition in the cage-cluster borides $RB_{12}$ (R- Ho, Er, Tm, Yb, Lu, Zr).

Gennady Komandin[a], Elena Zhukova[b], Boris Gorshunov[b], Alexander Melentyev[b], Ludmila Alyabyeva[b], Andrey Azarevich[a,b], Andrey Muratov[c], Yurii Aleshchenko[c], Nikolay Sluchanko[a,*]

[a]*Prokhorov General Physics Institute of RAS, 119991 Moscow, Vavilov str.,38, Russia*
[b]*Laboratory of Terahertz spectroscopy, Center for Photonics and 2D Materials, Moscow Institute of Physics and Technology (national Research University), Dolgoprudny, Moscow Region, Russia*
[c]*Lebedev Physical Institute of RAS, 119991 Moscow, Leninskiy Av. 53, Russia*

*\*e-mail: nes@lt.gpi.ru*

Keywords: cooperative Jahn-Teller instability, optical conductivity, quantum criticality, metal-insulator transition

**Abstract**
A thorough study of the wide-range (40-35000 $cm^{-1}$) dynamic conductivity spectra of the rare-earth (RE) dodecaborides $RB_{12}$ (*R*- Ho, Er and Tm) and $Tm_{1-x}Yb_xB_{12}$ substitutional solid solutions was carried out at room temperature. Both the Drude-type components and overdamped excitations have been separated and analyzed. An additional absorption band observed above 200 $cm^{-1}$ in these $RB_{12}$ with magnetic RE ions is attributed to the cooperative Jahn-Teller dynamics of the $B_{12}$ complexes, which depends crucially on the RE-ion cage space and is compared with the same effect found in the non-magnetic $LuB_{12}$ and $ZrB_{12}$. It was shown that non-equilibrium (hot) electrons participating in the formation of the collective JT modes dominate in charge transport, and portion of Drude-type carriers changes by 20-40% in these compounds with unstable boron lattice. Strong renormalization of the infrared response is observed in $Tm_{1-x}Yb_xB_{12}$ solid solutions with metal-insulator transition (MIT) and is discussed in terms of *localized* collective modes caused by Yb-ion valence instability. We demonstrate that even at room temperature the MIT is accompanied with simultaneous decrease in concentration of Drude-type electrons and redistribution of carriers to *localized* JT collective modes.

## 1. Introduction

Rare-earth (RE) and transition metal (TM) dodecaborides $RB_{12}$ with a "cage-glass" structure [1] attract considerable attention due to a unique combination of their physical properties, including high melting temperature, microhardness, high chemical stability, etc. In the family, there are several RE antiferromagnetic (AF) metals, whose Néel temperature decreases monotonically from $T_N \approx 22$ K ($TbB_{12}$) to $T_N \approx 3.2$ K ($TmB_{12}$), while the conduction band remains similar, consisting of 5*d* (*R*) and 2*p* (B) atomic orbitals and changing only the filling of the 4*f* shell of the RE ion ($8 \leq n_{4f} \leq 14$) [1-3]. The principal magnetic interaction, which couples the moments of unfilled 4*f* orbitals of $R^{3+}$ ions located at large distances (~5.3 Å) is the Ruderman-Kittel-Kasuya-Yosida (RKKY) indirect exchange. The long-range and oscillatory character of this coupling in the presence of other interactions (crystal-electric field, magnetoelastic coupling, manybody effects, etc.) may lead to a competition among the interionic interactions, resulting in frustration and emergence of complicated magnetic structures. In addition, the cooperative Jahn–Teller (JT) instability of $B_{12}$ clusters (ferrodistortive effect) is developed in these AF metals with simple face centered cubic (fcc) crystal structure [4, 5] (see inset in Fig. 1a), being among the main factors responsible for the formation of the *dynamic charge stripes* along ⟨110⟩ directions in the fcc lattice of $RB_{12}$, destroying the RKKY exchange between nearest RE magnetic ions and producing complicated magnetic diagrams in form of the Maltese Cross ($HoB_{12}$ [6], $Tm_{1-x}Yb_xB_{12}$ [7-8]) and butterfly-type (ErB$_{12}$ [9], DyB$_{12}$ [10]) with a number of various magnetic phases and phase boundaries [6-10].

Furthermore, the filling of 4$f$-shell between $n_{4f}$ = 12 (TmB$_{12}$) and $n_{4f}$ ~ 13 (YbB$_{12}$) in the RE dodecaborides leads to dramatic changes both in the magnetic and charge transport characteristics [11-13], demonstrating the transition from an AF metal to a paramagnetic narrow-gap semiconductor with the mixed valence $v \approx$ 2.9-2.95 of Yb ions [11, 14-15]. The metal-insulator transition (MIT) leads to a considerable growth of resistivity at helium temperatures, from 4 $\mu\Omega\times$cm in TmB$_{12}$ to about 10 $\Omega\times$cm in YbB$_{12}$, changing its behavior from metallic to semiconductor-like [1-2]. In spite of intensive investigations, the nature of the nonmagnetic semiconducting state in YbB$_{12}$ remains a subject of active debates [1, 16-26]. Recently, a special attention was paid to the mysterious insulating ground state in YbB$_{12}$ with a metallic Fermi surface, which was discussed in terms of extraordinary gapless charge-neutral fermionic excitations [20-24], but the finding of these Majorana fermions was refuted in the detailed heat capacity and Hall effect studies [27-29].

The two non-magnetic high borides ZrB$_{12}$ and LuB$_{12}$ in the family are believed to be model superconductors for various classes of high-$T_c$ (HTSC) compounds, including HTSC cuprates with collective dynamics of O$_6$ octahedrons centered on Cu-ions and $R$H$_n$ polyhydrides with $T_c$ ~250–270 K (see, for example, [30-32]). Indeed, two-gap superconductivity has been established in ZrB$_{12}$ in various experiments (see [33] for a recent review). In addition, the formation of grids of dynamic charge stripes and sub-structural charge density waves (s-CDWs) has been detected in [33-34], alongside the pseudo-gap state observed in ZrB$_{12}$ in photoemission spectroscopy experiments [35]. These findings are considered usually as the fingerprints of the unconventional HTSC, suggesting common features of unconventional superconductivity in ZrB$_{12}$, and at least, in cuprates and Fe-based pnictides. A new *vibron-plasmon-phonon mechanism* of superconducting pairing was proposed in [33], which may be common for various classes of HTSCs, including cage-cluster polyhydrides (La,Y)H$_n$ (n=10) with fcc crystal structure (space group $Fm\bar{3}m$) and the highest $T_c \geq$ 250 K discovered at pressures up to 170 GPa.

When discussing the mechanisms, responsible for the fcc lattice and electron instabilities in RE and TM dodecaborides, it is worth noting that the cooperative JT lability of B$_{12}$ clusters (ferrodistortive effect) should be considered as the source of (i) small static distortions of fcc lattice, which have been observed in LuB$_{12}$ [4-5, 36], HoB$_{12}$ and ErB$_{12}$, [5, 37], YbB$_{12}$ [27, 5, 37] and ZrB$_{12}$ [33], in conjunction with (ii) the strong dynamic JT effect, that develops within the rigid boron cage. The dynamic JT instability in compounds of $R$B$_{12}$ series was observed in the optical conductivity spectra whose anomalies were discussed in terms of overdamped oscillators discovered in LuB$_{12}$ [38], Tm$_{0.19}$Yb$_{0.81}$B$_{12}$ [39, 40] and ZrB$_{12}$ [41]. In particular, the dynamic conductivity measurements which were carried out on LuB$_{12}$ with various $^{10}$B and $^{11}$B isotope content [38] and on ZrB$_{12}$ [41], allow one to conclude that 60–80% of conduction electrons are involved in formation of collective JT excitations (hot charge carriers). Keeping in mind that the collective modes are a consequence of the lability of B$_{12}$ clusters [4-5], and in the case of Yb-based $R$B$_{12}$ compounds the JT lattice instability is accompanied with valence instability of Yb-ions, it looks promising to investigate the wide-range optical spectra of $R$B$_{12}$ (R-Ho, Er, Tm) and Tm$_{1-x}$Yb$_x$B$_{12}$ substitutional solid solutions. In this study, we will separate and analyze in detail optical conductivity of these AF metals and narrow-gap semiconductors in comparison with similar room temperature characteristics of the non-magnetic metals LuB$_{12}$ and ZrB$_{12}$. Thus, the goal of the present study is to establish the role of collective modes observed in the optical spectra of $R$B$_{12}$ and their various regimes in the implementation of MIT. It will be shown below that non-equilibrium (hot) electrons participating in the formation of the collective JT modes dominate in charge transport, and the share of the rest, carriers of the Drude type, changes by 20-40% in these compounds with unstable boron lattice. Strong renormalization of the infrared response is observed in Tm$_{1-x}$Yb$_x$B$_{12}$ solid solutions with metal-insulator transition, and discussed in terms of localized collective modes caused by Yb-ion valence instability. We demonstrate that MIT is accompanied in Tm$_{1-x}$Yb$_x$B$_{12}$ series with simultaneous decrease in the concentration of Drude-type electrons and redistribution of carriers between delocalized and localized JT collective modes.

## 2. Experimental details

High-quality HoB$_{12}$, ErB$_{12}$ and Tm$_{1-x}$Yb$_x$B$_{12}$ single crystals with the ytterbium concentrations $x$(Yb)=0; 0.25; 0.57; 0.6 and 1 and with the natural boron isotope content (which is a mixture of isotopes $^{10}$B (~19%) and $^{11}$B (~81%)) were grown according to the procedure described in detail in [42]. All samples were characterized carefully by the X-ray diffraction, chemical microanalysis and charge transport measurements. For infrared (IR) measurements, round ≈ 5 mm in diameter samples were prepared by polishing (to within ±1 μm) their surface that was subsequently etched in the boiled HNO$_3$+H$_2$O solution to avoid extra phases and structural distortions. All measurements were performed at room temperature. In the range of frequencies $v$=40-8000 cm$^{-1}$, IR reflectivity spectra $R(v)$ were measured using Vertex 80V Fourier-transform spectrometer with the gold films deposited on a glass substrate used as reference mirrors. With the J.A. Woollam V-VASE ellipsometer, real and imaginary parts of dielectric permittivity and AC conductivity of the samples were directly determined in the frequency interval 3700 – 35000 cm$^{-1}$ with resolution 50 cm$^{-1}$. From the ellipsometry data, reflection coefficients have been calculated using standard Fresnel equations and then merged with the measured IR reflectivity spectra. We used reflectivity data from [43-44] to extend the $R(v)$ spectra up to 400000 cm$^{-1}$; this allowed accurate processing of the spectra, as described below. In addition, a direct current (DC) resistivity and Hall effect of the same samples were measured using five-terminal scheme.

## 3. Experimental results

Fig. 1 shows a broad-band reflectivity spectra $R(v)$ of (a) HoB$_{12}$ and ErB$_{12}$ and (b) Tm$_{1-x}$Yb$_x$B$_{12}$ crystals (dots) of several selected ytterbium concentrations. Spectra obtained of HoB$_{12}$, ErB$_{12}$ and TmB$_{12}$ have typical metallic shape [45-46]: there is a plasma edge (minimum) at a frequency of $\frac{v_{pl}}{\sqrt{\varepsilon_\infty}} \approx$ 14000 cm$^{-1}$ (here $\varepsilon_\infty$ is the high-frequency dielectric permittivity and $v_{pl}$ is the plasma frequency) and close to 100% reflectivity values at lower frequencies. When going to higher contents of ytterbium, the plasma edge practically does not change showing only slight decrease in depth. At the same time, there are significant changes in the spectra in the range 100-10000 cm$^{-1}$. Here, the reflectivity values are getting lower and, besides, broad and smooth features gradually emerge. To extract the spectra of AC conductivity of the samples, we performed a least-square fitting analysis of the reflectivity data. Free-carrier conductivity of the samples was taken into account by introducing the Drude conductivity model that provides an expression for frequency-dependent complex conductivity $\sigma_D(v)$ [45-46]

$$\sigma_D(v) = \frac{\sigma_{DC}}{1 - iv/\gamma_D} \qquad (1),$$

where $\sigma_{DC}$ is the direct current conductivity and $\gamma_D$ is the charge-carrier scattering rate. To describe broad features in the spectra, we used *minimal* set of Lorentzian terms given as

$$\sigma(v) = \frac{0.5 f v}{v\gamma + i(v_0^2 - v^2)} \qquad (2)$$

with $v_0$ being the resonance frequency, $f = \Delta\varepsilon v_0^2$ - the oscillator strength, $\Delta\varepsilon$ - the dielectric contribution (strength) and $\gamma$ - the damping constant. Examples of the results of least-square fitting are shown by lines in Fig. 2a-2h where the obtained spectra of the IR conductivity $\sigma(v)$ are shown together with the DC data (marked by red arrows). It turned out, that for all RE dodecaborides with the natural boron isotope content it is sufficient to involve five terms to describe the measured reflectivity spectra: the Drude term (D) and four Lorentzian terms (L$_1$-L$_4$); note that the Lorentzian contributions L$_1$-L$_2$ were found to be significantly more intensive compared to L$_3$-L$_4$ terms (Fig. 2).

Similar to the conclusions drawn in [28, 38-41, 47-49], we assume that the origin of the observed collective excitations is related to the JT instability of the B$_6$ and B$_{12}$ complexes of natural boron in $R$B$_6$ and $R$B$_{12}$, respectively, leading to the emergence of both small *static* and strong *cooperative dynamic* JT effects [5]. We suggest that this kind of non-adiabatic mechanism, which launches both the collective overdamped modes and the electronic instability, provides the most natural interpretation of the anomalies observed in the $\sigma(\nu)$ spectra of $RB_{12}$ [1, 5]. Taking into account that the JT instability of rigid boron cage is among the key factors responsible for the emergence of the collective modes in $RB_{12}$, it is natural to attribute the variation of the characteristics of both Drude contribution (1) and Lorentzians (2) to changes in the cage space parameter $2(d_{R-B} - r_{RE})$. Here $d_{R-B}$ and $r_{RE}$ are the distance between RE ion and boron atom (accurately detected in [5, 36-37, 50]), and the ionic radius of RE ion [51], respectively. The cage space may be considered, as the characteristics of motion of the RE atoms filling oversized lattice cages of boron atoms. Along the $RB_{12}$ series, the enhancement of free space for RE-ion motion is expected being attributed to the lanthanide contraction. Figures 3 and 4 show the dependence on the cage space $2(d_{R-B} - r_{RE})$ of the room temperature parameters of the Drude model (1) (see Fig. 3a-3c) and of the absorption peaks $L_1$-$L_4$ (see Fig. 4a–4d).

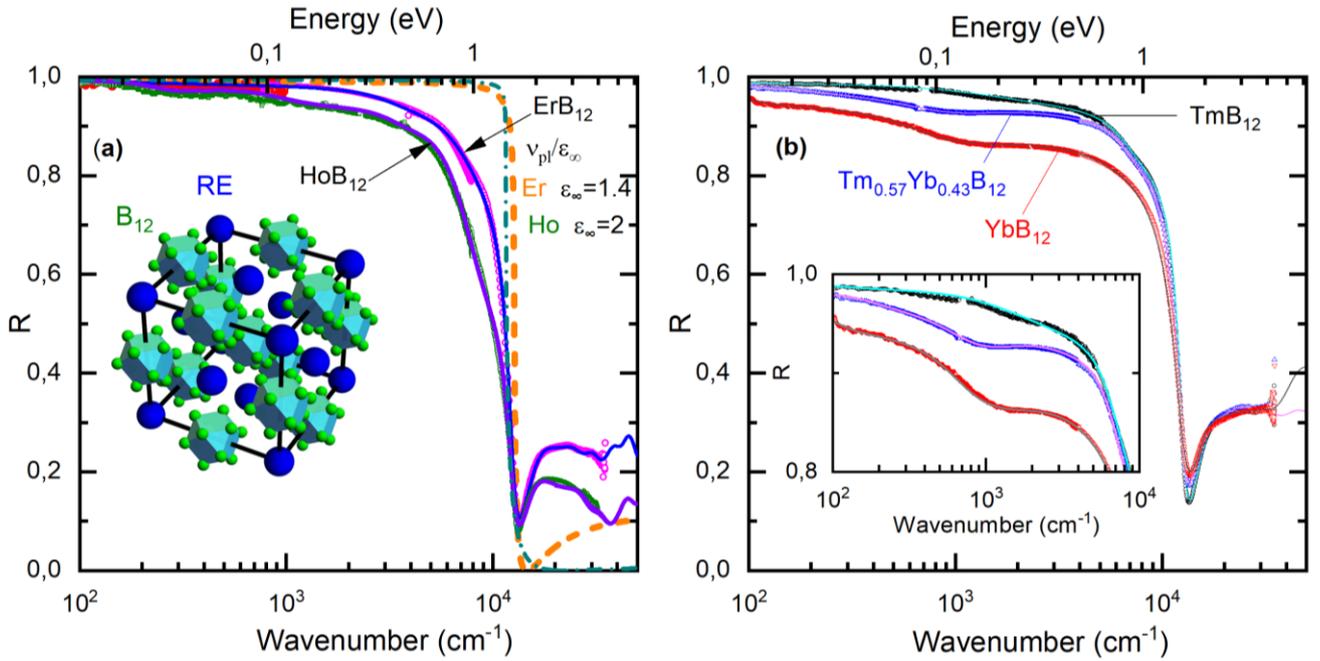

**Fig. 1.** (Color online). Room temperature reflection coefficient spectra of $HoB_{12}$ and $ErB_{12}$ (a) and $Tm_{1-x}Yb_xB_{12}$ (b) single crystals. The insets in panels (a) and (b) represent the crystal structure of $RB_{12}$ and the expanded low-frequency reflectivity spectra of $Tm_{1-x}Yb_xB_{12}$, respectively. Dots show experimental data obtained using Fourier-transform spectrometer and ellipsometer. Lines above 35000 cm$^{-1}$ correspond to high-frequency reflectivity data from [43-44]. Solid lines show the results of fitting the spectra using the Drude term (1) for the free charge carrier response and Lorentzians (2), responsible for absorption resonances. Dashed lines in panel (a) show best fits of the spectra that have been plotted using the Drude conductivity term alone with $\sigma_{DC}$ and $\gamma_D$ presented in Figures below.

It is seen in Fig. 3, that even at room temperature both the DC conductivity (Fig. 3a) and plasma frequency (Fig. 3c) decrease strongly during the metal-insulator transition in the $Tm_{1-x}Yb_xB_{12}$ series (MIT interval is marked by light-green color in Figs. 3-5), and both these parameters reach their minimum values in the so-called Kondo-insulator (KI) $YbB_{12}$. The relaxation rate $\gamma_D$ of Drude carriers changes only slightly between $ErB_{12}$ and $LuB_{12}$, but it demonstrates a noticeable peak near the quantum critical point at $x_c \sim 0.25$ (QCP in Figs. 3-5), where the antiferromagnetic instability develops in $Tm_{1-x}Yb_xB_{12}$, accompanied by a zeroing of the Néel temperature [2, 8, 50]. Note, that quantum critical behavior with QCP at $x_c \sim 0.25$ was detected previously in accurate x-ray diffraction studies of $Tm_{1-x}Yb_xB_{12}$ performed at room temperature [50]. In [52, 53], it was predicted that the arising antiferromagnetic instability near the QCP at $T_N=0$ should substantially modify the characteristics of strongly correlated electron systems in a wide temperature range, including room temperature, so, we may conclude in favor of the room temperature quantum criticality scenario predicted in [52-53].

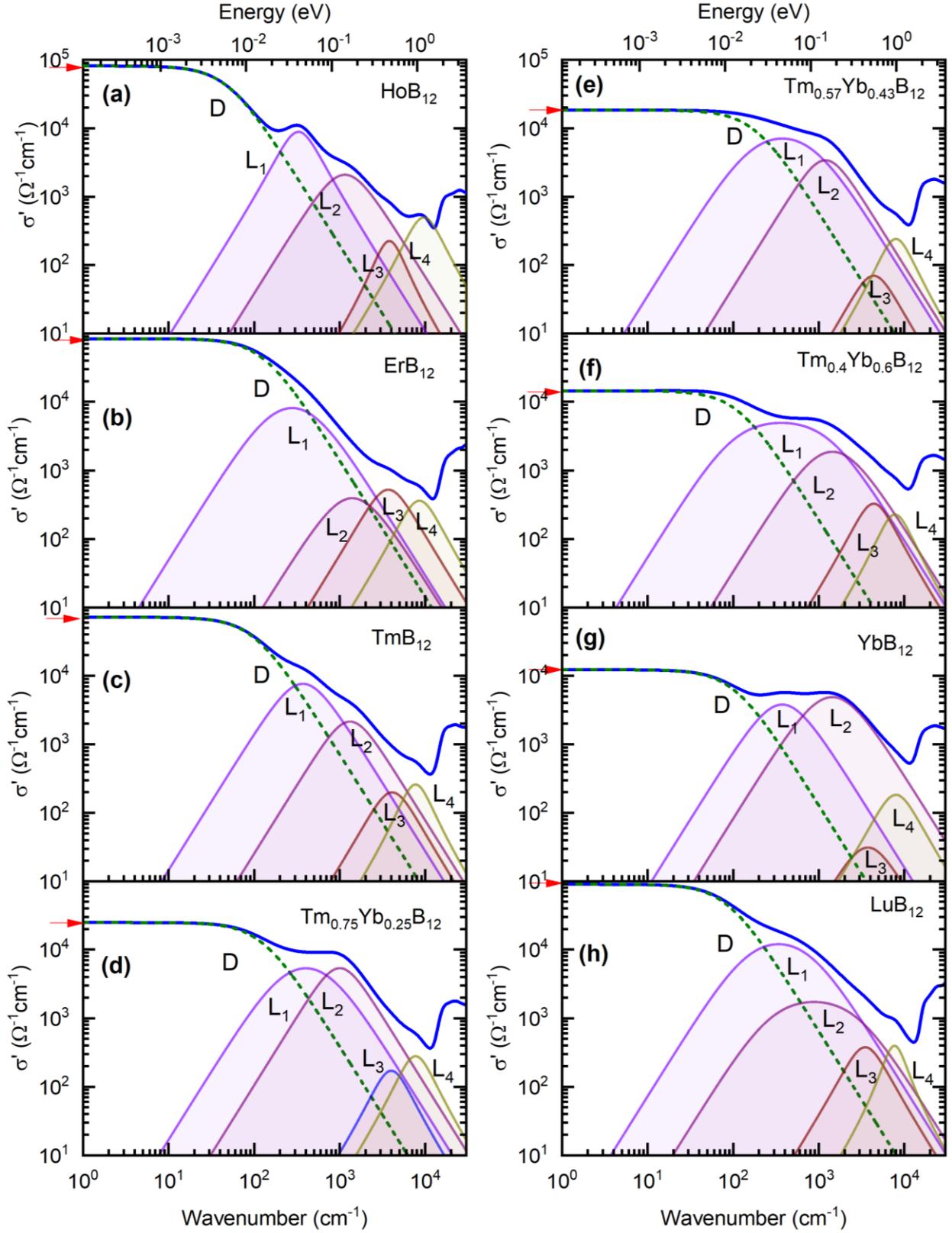

**Fig. 2.** (Color online). Room temperature spectrum of real part of conductivity of (a) HoB$_{12}$, (b) ErB$_{12}$, (c) TmB$_{12}$ (d-g), Tm$_{1-x}$Yb$_x$B$_{12}$ (x= 0.25, 0.43, 0.6 and 1) and (h) LuB$_{12}$ (data from [38]) crystals (solid lines), obtained by least square fitting of the reflection coefficient spectra using Drude conductivity (1) and Lorentzian (2) terms, as described in the text. The Drude (D) and Lorentzian (L$_1$-L$_4$) contributions are shown separately. Red arrows at the left of conductivity axes mark the $\sigma_{DC}$ values.

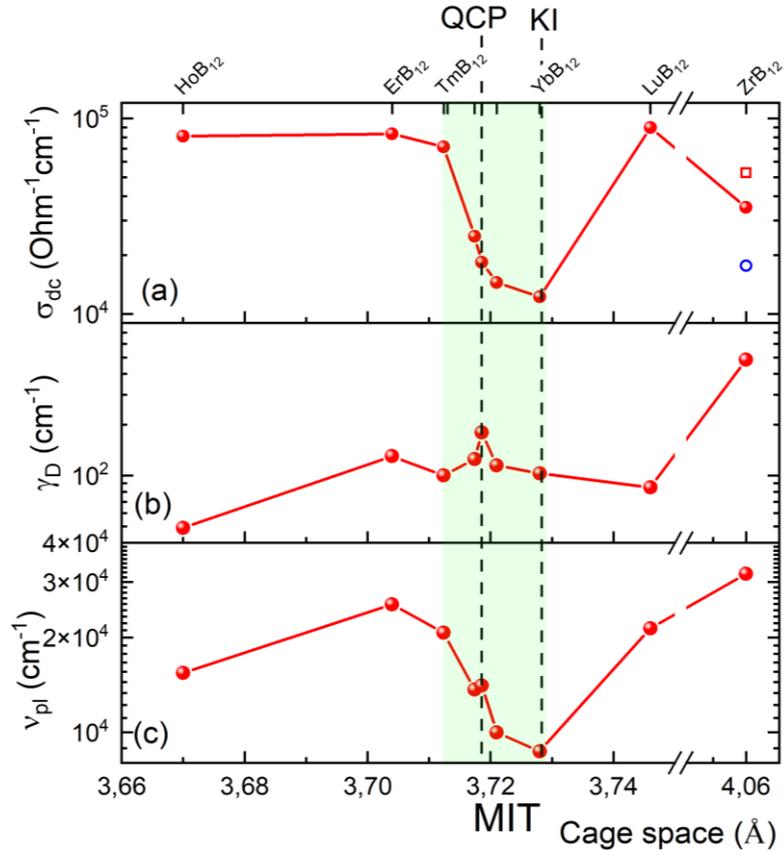

**Fig. 3.** (Color online) (a) Room temperature dependences on the cage space $2(d_{R-B} - r_{RE})$ (see text) of charge carriers' characteristics related to the Drude conductivity in the $R$B$_{12}$ series: the DC conductivity $\sigma_{DC}$, the charge carriers' scattering rate $\gamma_D$ and the plasma frequency $\nu_{pl}$. Light green area marks the interval where the metal-insulator transition (MIT) occurs. QCP and KI denote the antiferromagnetic quantum critical point (Néel temperature $T_N \sim 0$ at $x$(Yb) $\approx$ 0.25 in Tm$_{1-x}$Yb$_x$B$_{12}$) and the so-called Kondo-insulator YbB$_{12}$, respectively.

The position of the collective modes L$_1$-L$_4$ in the conductivity spectra changes moderately in the $R$B$_{12}$ series, except for a significant anomaly near the QCP for the L$_2$ line (Fig. 4a). Anomalies at $x_c \sim$ 0.25 are observed also in the dependences of relaxation rate $\gamma$ and dielectric contribution $\Delta\varepsilon$ of L$_2$ and L$_3$ modes (Figs. 4b-4c). The intervals of change of the characteristics $\nu_0$, $\gamma$, and $\Delta\varepsilon$ of these L$_1$-L$_4$ absorption lines along the $R$B$_{12}$ series practically do not overlap, whereas for the oscillator strength $f\mathrm{osc} = \Delta\varepsilon\nu_0^2$ the situation is completely different. Indeed, $f_{osc}$ demonstrates significant variation during the metal-insulator transition (MIT, see Fig. 4), and particularly strong increase in $f_{osc}$ is observed for YbB$_{12}$, where the $f_{osc}$ of L$_2$ line reaches its maximum values in the so-called Kondo-insulator regime (YbB$_{12}$) (Fig. 4d).

Note again, that the observed absorption bands (L$_1$-L$_4$ in Figs. 2a-2h) have rather unusual characteristics. Firstly, these collective modes are clearly visible in both the reflectivity and conductivity spectra, as they are not completely screened by charge carriers. Secondly, L$_1$ and L$_2$ demonstrate rather large values of the oscillator strengths $f_{1,2} \sim (0.1-1.1) \times 10^9$ cm$^{-2}$, the dielectric contributions $\Delta\varepsilon_{1,2} \sim$ 100 –5000 and the damping constants $\gamma_j/\nu_{0j} \sim 0.5 - 5$ (see Fig. 4). This implies that indeed the relevant absorption mechanisms cannot be associated with usual interband transitions.

### 4. Discussion

When discussing the singularities in the $\sigma(\omega)$ spectra of $R$B$_{12}$ (Fig. 2), it is worth noting, that a broad mid-infrared (mIR) peak was observed in YbB$_{12}$ at ~250 meV [43,54-56]. It was demonstrated

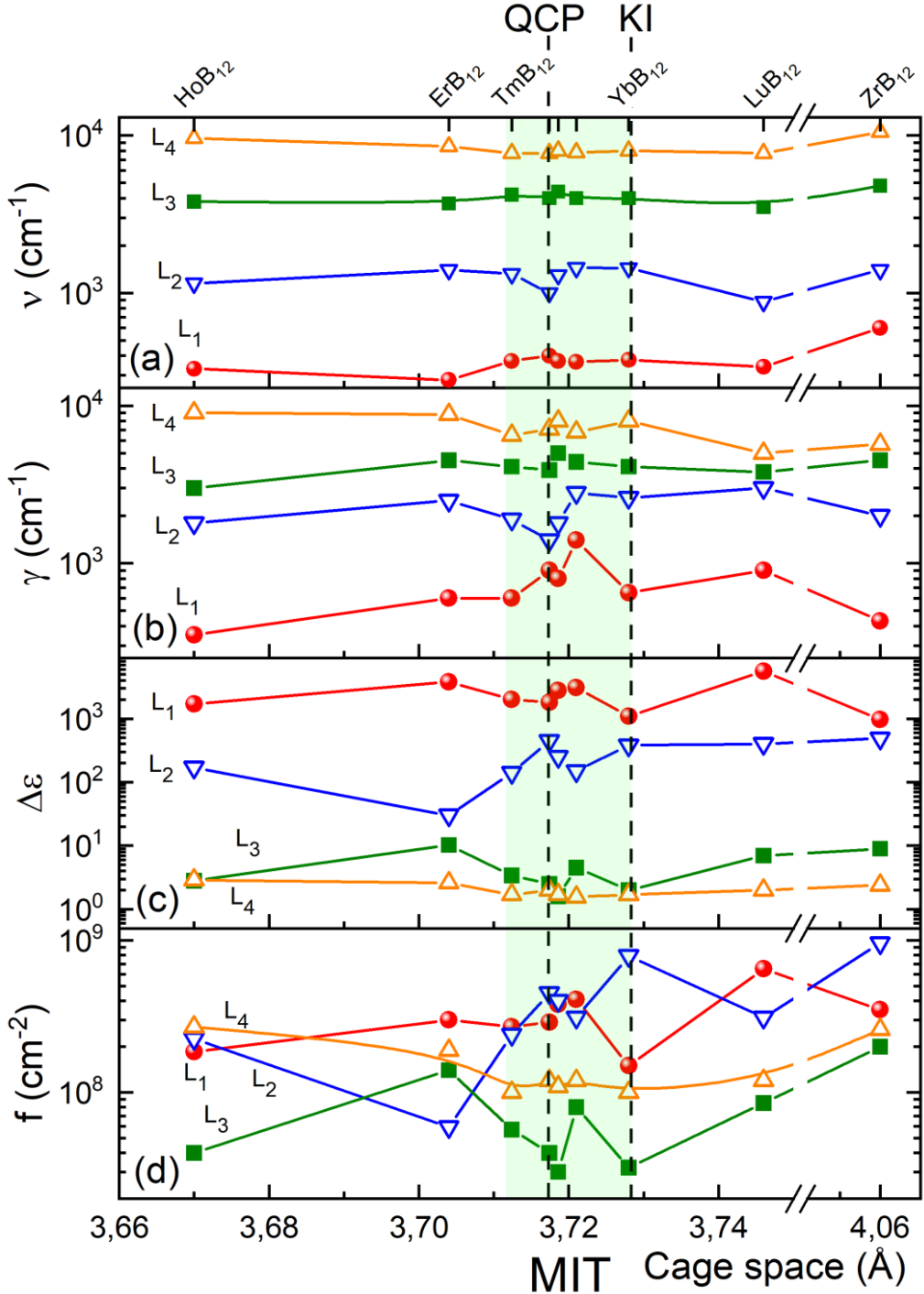

**Fig. 4.** (Color online). Room temperature dependences on the cage space $2(d_{R-B} - r_{RE})$ (see text) of the parameters of absorption bands $L_1$-$L_4$ in $R$B$_{12}$: (a) the resonance frequency $\nu_0$, (b) the damping constant $\gamma$, (c) the dielectric contribution $\Delta\varepsilon$, and (d) the oscillator strength $f\,\text{osc} = \Delta\varepsilon\nu_0^2$. Light green area marks the interval where the metal-insulator transition (MIT) occurs. QCP and KI denote the antiferromagnetic quantum critical point (Néel temperature $T_N \sim 0$ at $x(\text{Yb}) \approx 0.25$ in Tm$_{1-x}$Yb$_x$B$_{12}$) and the so-called Kondo-insulator YbB$_{12}$, respectively.

[55-56] that the peak position depends dramatically on the Lu concentration, decreasing monotonically down to 120 meV for $x = 7/8$ in Yb$_{1-x}$Lu$_x$B$_{12}$. In addition to the gradual shift of the mIR peak towards lower energy with increasing of Lu content, the displacement of the anomaly towards higher energy with decreasing temperature has also been observed [55-56]. Taking into account, that (i) a similar mid-

IR anomaly has been detected at room temperature for the nonmagnetic reference compound LuB$_{12}$ [38, 57] (see Fig. 2h), and (ii) the extrapolation of the band width and position in the series Yb$_{1-x}$Lu$_x$B$_{12}$ to $x$ = 1 gives the same value of ~800 cm$^{-1}$ (~100 meV) as was measured in LuB$_{12}$ [38,57], it is natural to conclude that the nature of these collective modes in YbB$_{12}$ and LuB$_{12}$ is common. Thus, the origin of the mIR peaks in Yb$_{1-x}$Lu$_x$B$_{12}$ [43,54-57] and in other RE dodecaborides (Fig. 2) can be attributed to the collective JT dynamics of the boron clusters B$_{12}$ (ferrodistortive effect, see [5] for more detail), which is additionally renormalized by the fast charge and spin fluctuations on the Yb ions in the case of Yb-based compounds.

    It should be pointed out that the situation is much more complicated in the dodecaborides with Yb ions, where, in addition to the JT instability of the boron cage, the instability of Yb 4$f$ –electron configuration appears, providing one more mechanism of the charge and spin fluctuations in $R$B$_{12}$. Emergence of Yb–Yb pairs during cooling can be probably considered as the main factor responsible for the charge- and spin-gap formation in YbB$_{12}$, and the intra-gap many-body states are characterized by the localization radius ~5 Å [54, 58, 59], which is close to the Yb–Yb distance (~5.3 Å) in the fcc lattice [27]. Taking into account that the indirect gap opening is a phonon-assisted process [54, 60] and that there is coupling between the intensities of the anomalous acoustic phonons and the M1 and M2 magnetic excitations in YbB$_{1212}$ [61], it can be concluded that the Yb dimers are vibrationally coupled complexes.

To clarify the nature of weak localization of conduction electrons observed in $R$B$_{12}$ at room temperature, the characteristics of Drude term (D) and Lorentzians (L$_1$-L$_4$, see Figs. 2a-2h) were used to estimate the normalized concentration of (i) the Drude-type electrons $n_D m_0/m*$ and (ii) hot charge carriers $n_{peak} m_0/m*$ involved in the cooperative JT dynamics ($m*$ and $m_0$ are effective mass and mass of free electron, respectively). Specifically, using the relations for charge carriers plasma frequency $v_{pl} = [n_D e^2/\pi m^*]^{1/2}$ ($e$ is free electron charge) and sum of oscillator strengths of Lorentzians $f = \sum f_j = \sum \Delta\varepsilon_j v_{0j}^2 = n_{peak} e^2 (\pi m)^{-1}$, the ratios $n_D m_0/m*$ and $n_{peak} m_0/m*$ were obtained and the sum $n_{opt} m_0/m* = (n_D + n_{peak}) m_0/m*$ was compared with the Hall concentration $n_H$ of charge carriers (Fig. 5a). It is evident from Fig. 5a, that the fraction of the non-equilibrium (hot) electrons $n_{peak}$ exceeds the concentration of Drude-type charge carriers $n_D$ for all compounds in the $R$B$_{12}$ series. The value is changing by 50-90% of the total amount $n_{opt}$, while the parameter $n_{peak} m_0/m*$ is practically unchanged up to the QCP. Above $x_c$=0.25 within the MIT interval, the step-like increase of $n_{peak} m_0/m*$ is observed, and at least in the range 0.25< $x$ ≤1 for Tm$_{1-x}$Yb$_x$B$_{12}$, the normalized "optical" concentration $n_{opt} m_0/m*$ exceeds significantly the $n_H$ values. Fig. 5b shows a rough estimate of the charge carriers effective mass $m*/m_0 = n_{opt}/n_H$, which is compared with the cyclotron mass detected in the de Haas-van Alphen (dHvA) magnetization measurements of LuB$_{12}$ [62], HoB$_{12}$, ErB$_{12}$, TmB$_{12}$ [63] and ZrB$_{12}$ [64]. We can conclude that: (i) the charge transport mass $m*/m_0$ estimated here at $T \approx$300 K for metals LuB$_{12}$, HoB$_{12}$, ErB$_{12}$, TmB$_{12}$ and ZrB$_{12}$ is very close to cyclotron mass values obtained in the quantum oscillation dHvA experiments [62-64], performed at liquid helium temperatures; (ii) within the interval of MIT (Tm$_{1-x}$Yb$_x$B$_{12}$ with 0.25< $x$ ≤1), the average effective mass increases vastly (nearly twice) suggesting the tendency towards localization of charge carriers observed even at room temperature (Fig. 5b). Fig. 5c shows the dependences of mobility of charge carriers – the Hall mobility $\mu_H= R_H \sigma$ ($R_H$ is the Hall coefficient) and the mobility obtained for the Drude component $\mu_D =e/(2\pi\gamma_D m^*)$ in the $R$B$_{12}$ series. It is worth noting that the $\mu_D$ is much higher than $\mu_H$, which should be considered as the average mobility value for all types of conduction electrons. Moreover, a strong (~3 times) *decrease* in room-temperature Hall mobility between TmB$_{12}$ and YbB$_{12}$ is accompanied by a significant (~2.5 times) *increase* in $\mu_D$ (see Fig. 5c). We suggest that during the MIT, the localization of non-equilibrium (hot) electrons, which is accompanied by strong freezing out of the Drude-component (Fig. 5a), reduces significantly the electron-electron scattering between these free ($n_D$) and hot ($n_{peak}$) components, resulting in the drastically different dependences of the $\mu_H$ and $\mu_D$ on the cage space, see Fig. 5c.

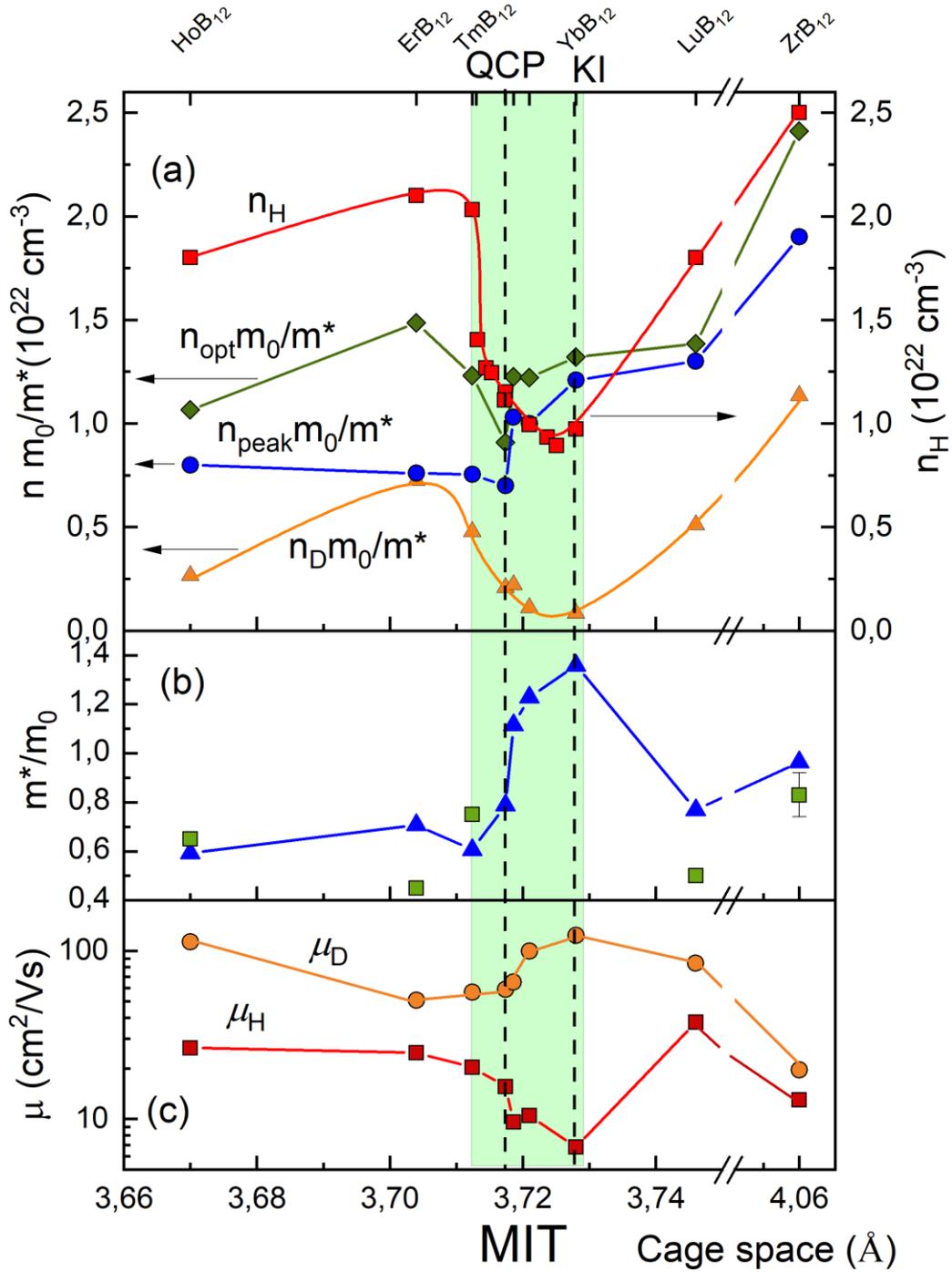

**Fig. 5.** (Color online) Room temperature dependences on the cage space $2(d_{R-B} - r_{RE})$ (see text) of (a) the normalized concentrations of the Drude-type charge carriers $n_D m_0/m^*$ and hot electrons $n_{peak} m_0/m^*$, their sum $n_{opt} m_0/m^* = (n_D + n_{peak}) m_0/m^*$, and the Hall concentration $n_H$. Panel (b) presents a rough estimation of the effective mass $m^*/m_0 \approx n_{opt}/n_H$ in the $R$B$_{12}$ series in comparison with the cyclotron masses (rectangular green symbols) deduced from the dHvA experiments [62-64]. Panel (c) shows the dependences of Hall $\mu_H = R_H/\rho$ and Drude $\mu_D = e/(2\pi\gamma_D m^*)$ charge carriers' mobility in the $R$B$_{12}$ series. Light-green area marks the interval where the metal-insulator transition (MIT) occurs. QCP and KI denote the antiferromagnetic quantum critical point (Néel temperature $T_N \sim 0$ at $x(Yb) \approx 0.25$ in Tm$_{1-x}$Yb$_x$B$_{12}$) and the so-called Kondo-insulator YbB$_{12}$, respectively.

It is reasonable to suggest that some of the overdamped excitations identified in the room-temperature conductivity spectra of $Tm_{1-x}Yb_xB_{12}$ should be regarded as *localized modes* and these excitations do not affect the charge transport at liquid helium temperatures. Such kind of absorption bands were discovered recently in the low temperature optical conductivity spectra of the strongly correlated narrow-gap semiconductor $Tm_{0.19}Yb_{0.81}B_{12}$ with MIT [40]. Indeed, the broad absorption band in the conductivity spectrum centered at about 2000 cm$^{-1}$ was observed for $Tm_{0.19}Yb_{0.81}B_{12}$, and it changes only slightly when cooling from 300 K to 10 K [40]. In [40], the origin of this band was associated with the vibrationally coupled localized states located well below the Fermi level in the conduction band. The position, width and dielectric contribution of these absorption peaks in $Tm_{1-x}Yb_xB_{12}$ (see Fig. 2 and [40]) are similar, arguing in favor of the proposed interpretation.

Based on the above it is worth emphasizing, that (i) the localization mechanism underlying the formation of the charge and spin gaps and of heavy fermions in the vicinity of Yb ions in $YbB_{12}$ is different from the Kondo-insulator scenario; (ii) the localization radius for these intra-gap many-body states turns out to be nearly equal to the Yb–Yb distance (~5 Å [27]), being much smaller than that of the "Kondo cloud" (> 20 Å). Taking into account, that the anisotropic Kondo Hamiltonian is a special case of the "spin-boson" Hamiltonian, which describes the properties of a dissipative two-level system [65], it is natural to argue [66,1], that the origin of spin and charge fluctuations in $YbB_{12}$ is related to the coupled vibrations of the heavy particles (RE ions) between states in the double-well potential, and that the Kondo physics is invalid in this case. Moreover, within the framework of the common scenario of the cooperative JT lattice instability developed in the $RB_{12}$ family [1, 5], the low temperature symmetry breaking was identified in $YbB_{12}$ [28]. It has been shown [28] that this archetypal strongly correlated system with a metal-insulator transition to a mysterious dielectric ground state with a metallic Fermi surface [20-24] is actually a *heterogeneous* compound in the regime of nanoscale *electronic phase separation*. The authors argued in favor of appearance upon cooling of dynamic charge stripes and the filamentary electronic structure, which penetrates the semiconducting matrix of $YbB_{12}$ and includes also vibrationally coupled Yb-Yb pairs [27-28]. The arrangement of both the Yb-pairs and the charge stripes leads to the *spontaneous selection of a special direction* in the formally cubic crystal with the patterns of these fluctuating charges being extremely sensitive to the strength and direction of external magnetic field. When discussing the changes in the characteristics of the collective modes $L_1$-$L_4$ (Fig. 4), it may be suggested that the increase in the oscillator strength of the $L_2$ band and the localization of electrons in the $L_2$ band need to be regarded as the main reason of MIT in the in $RB_{12}$ family. Low-temperature optical experiments on $Tm_{1-x}Yb_xB_{12}$ (0< $x$ ≤1) are in progress which will shed more light on the nature of both the room temperature weak localization phenomenon and the mechanism of the MIT underlying the formation of these localized manybody states.

**Conclusions.**

A detailed, wide-range (40-35000 cm$^{-1}$) study of the room temperature dynamic conductivity $\sigma(v)$ spectra of the rare-earth dodecaborides $RB_{12}$ with Ho, Er, Tm and Yb RE ions was carried out, and the results were analyzed and compared with $\sigma(v)$ contributions detected in the non-magnetic reference compounds $LuB_{12}$ and $ZrB_{12}$. Both the Drude-type term and several collective modes - overdamped excitations have been separated and characterized. An additional absorption band, which depends crucially on the $R$-ion cage space $2(d_{R-B} - r_{RE})$, was observed above 200 cm$^{-1}$ in these $RB_{12}$ with various $R$- ions and it is attributed to the cooperative Jahn-Teller dynamics of the $B_{12}$ complexes. Unusual room temperature anomalies of a number of spectral characteristics were established near quantum critical point (QCP) $x_c$~0.25 in the series of substitutional solid solutions $Tm_{1-x}Yb_xB_{12}$ resulting from the development of antiferromagnetic instability with the Néel temperature zeroing at $x_c$. It was shown that non-equilibrium (hot) electrons participating in the formation of the collective JT modes dominate in charge transport, and portion of Drude-type carriers changes by 20-40% in these compounds with unstable boron lattice. The metal-insulator transition (MIT) observed at room temperature in the $Tm_{1-x}Yb_xB_{12}$ series is accompanied by the emergence of weak localization of hot electrons in collective modes caused by the valence instability of Yb ions. In summary, the analysis developed here demonstrates that the Kondo-insulator scenario of MIT is invalid in YbB$_{12}$, and convincingly supports the electron-phonon mechanism of charge carrier localization.

ACKNOWLEDGEMENTS
The work was partly performed using the equipment of the Shared Facilities Center of Lebedev Physical Institute of RAS. Infrared experiments were supported by the Ministry of Science and Higher Education of the Russian Federation (FSMG-2025-0005). The analysis of FIR spectra performed by G. A. K. was supported by the Russian Science Foundation, project No. 25-79-30006. The authors are grateful to S.V. Demishev, V.V. Glushkov, N. B. Bolotina, S. Gabani and K. Flachbart for useful discussions.